\documentclass[12pt]{article} 
\usepackage{bbold}
\usepackage{cite}
\usepackage{graphicx}
\def \bea{\begin{eqnarray}} 
\def \beq{\begin{equation}}

\def \eea{\end{eqnarray}} 
\def \eeq{\end{equation}}

\def\lsim{\mathrel{\rlap{\lower3pt\hbox{$\sim$}}\raise2pt\hbox{$<$}}}
\def\gsim{\mathrel{\rlap{\lower3pt\hbox{$\sim$}}\raise2pt\hbox{$>$}}}

\begin{document} 
\begin{flushright}
TECHNION-PH-2016-09 \\
EFI 16-13 \\
August 2016 \\
\end{flushright} 
\centerline{\bf IMPROVING THE MEASUREMENT OF THE CKM PHASE $\phi_2 =\alpha$}
\centerline{\bf IN $B \to \pi \pi$ AND $B \to \rho \rho$ DECAYS}
\medskip
\centerline{Michael Gronau}
\centerline{\it Physics Department, Technion, Haifa 32000, Israel}
\medskip 
\centerline{Jonathan L. Rosner} 
\centerline{\it Enrico Fermi Institute and Department of Physics,
  University of Chicago} 
\centerline{\it Chicago, IL 60637, U.S.A.} 
\bigskip

\begin{quote}
CP-violating asymmetries in $B \to \pi \pi$ and $B \to \rho \rho$ decays can
help specify the weak phase $\phi_2 = \alpha$ of the Cabibbo-Kobayashi-%
Maskawa (CKM) matrix.  We discuss the impact of improved measurements of
these processes such as will be available in the near future, finding special
value in better measurement of the time-dependent CP violation parameter
$S_{00}$ in $B^0 \to \pi^0 \pi^0$ and $B^0 \to \rho^0 \rho^0$.  Reducing the
errors on $B \to \rho \rho$ measurements by a factor of two can potentially
lead to an error in $\phi_2 = \alpha$ just above $2^\circ$, at which level
the $\rho$ width and isospin-breaking corrections must be considered.
\end{quote}

\leftline{\qquad PACS codes: 12.15.Hh, 13.25.Hw, 14.40.Nd}
\bigskip

Precision measurements of the phases of weak charge-changing transitions,
as encoded in the Cabibbo-Kobayashi-Maskawa (CKM) matrix, are a potential
window to new physics if inconsistencies are uncovered.  The unitarity of
the CKM matrix may be expressed in terms of a triangle in the complex plane,
expressing the relation
\beq \label{eqn:ut}
V^*_{ub} V_{ud} + V^*_{cb} V_{cd} + V^*_{tb}V_{td} = 0~.
\eeq
In present fits to data the angles
of the triangle add up to $\pi$ within a few degrees, as illustrated in
Table \ref{tab:angles}.  Small differences between the fits of Refs.\
\cite{CKMfitter} and \cite{UTfit} may be ascribed to differing inputs and
statistical methods, and are indicative of present systematic uncertainties.

The weak phase $\beta = \phi_1$ is measured with fractional-degree
accuracy by CP asymmetries in such processes as $B^0 (\bar B^0) \to
J/\psi K_S$.  Individual measurements of the CKM phases $\alpha$
and $\gamma$ carry considerably larger uncertainties. The phase $\alpha$
can be extracted from isospin analyses of $B \to \pi \pi$ and $B \to \rho\rho$
decays \cite{Gronau:1990ka}.  For instance, the Babar collaboration
\cite{Lees:2012mma} has used $B \to \pi\pi$ to constrain this phase to a range
$71^\circ < \alpha < 109^\circ$ at a $1\sigma$ level, while Belle
\cite{Adachi:2013mae} obtained a weaker constraint. More precise determinations
of $\alpha$ have been obtained from analyses of longitudinally polarized $B \to
\rho \rho$, for which Babar \cite{Aubert:2009it} and Belle
\cite{Vanhoefer:2015ijw} find values of $\alpha$, $(92.4^{+6.0}_{-6.5})^\circ$ 
and $(93.7 \pm 10.6)^\circ$, respectively. A smaller uncertainty can be
obtained from $B \to \rho\rho$ analyses relying on the approximate validity of
SU(3) \cite{Beneke:2006rb}. Studying $B \to \pi \rho$ decays is more
complicated as a result of the non-identity of the final-state particles
\cite{Snyder:1993mx,Lees:2013nwa}.

Isospin analyses usually neglect a higher-order electroweak penguin amplitude 
\cite{Neubert:1998pt} and isospin-breaking effects.  Inclusion of the former
amplitude decreases the value of $\alpha$ determined in $B\to \pi\pi, \rho\rho$
by a calculable amount of $1.8^\circ$~\cite{Gronau:1998fn,Gronau:2005cz}.
Uncertainties at this same small level are
introduced by isospin-breaking corrections \cite{Gardner:1998gz,Gronau:2005pq}
and by a finite $\rho$ width effect \cite{Falk:2003uq}. 

In this note we concentrate on ways to improve the determination of $\alpha$
from $B \to \pi \pi$ and $B \to \rho \rho$ decays using isospin, by identifying
the major sources of statistical and systematic error.  We identify one
uncertainty as the large statistical error in the difference between
time-integrated rates for $B^0 \to \pi^0 \pi^0$ and $\bar B^0 \to \pi^0 \pi^0$,
encoded in the parameter $C_{00}$, and another in the parameter $S_{00}$
measured in time-dependent studies.  The uncertainty in the branching fraction
for $B^+ \to \pi^+ \pi^0$ could stand some improvement as well.  As has been
noted \cite{Vanhoefer:2015ksa}, measurement of time-dependent CP violation in
$B^0 (\bar B^0) \to \pi^0 \pi^0$ can help to reduce discrete ambiguities in
the determination of $\alpha$.  We find that $B \to \rho \rho$ decays are
subject to the same discrete ambiguity arising in the extraction of $\alpha$  
from $B \to \pi \pi$ decay.  The error in $B \to \rho \rho$ decays can be
reduced by improving measurements of the longitudinal branching ratios for
$B^0\to \rho^+\rho^-$ and $B^+ \to \rho^+\rho^0$, and especially by improving
measurement of the parameter $S_{00}$ describing time-dependent CP violation in
$B^0 \to \rho^0 \rho^0$.

\begin{table} 
\caption{Fits to angles of the unitarity triangle expressing the sum rule
(\ref{eqn:ut}) as quoted by CKMfitter \cite{CKMfitter} and UTfit \cite{UTfit}.
\label{tab:angles}}
\begin{center}
\begin{tabular}{c c c c} \hline \hline
  & $\alpha = \phi_2 = $ & $\beta = \phi_1 = $ & $\gamma = \phi_3 = $ \\
Fit & Arg(--$V^*_{tb}V_{td}/V^*_{ub}V_{ud}$) &
      Arg(--$V^*_{cb}V_{cd}/V^*_{tb}V_{td}$) &
      Arg(--$V^*_{ub}V_{ud}/V^*_{cb}V_{cd}$) \\ \hline
CKMfitter & $90.4^{+2.0}_{-1.0}$ & $22.62^{+0.44}_{-0.22}$ &
 $67.01^{+0.88}_{-1.99}$ \\
UTfit & $88.6 \pm 3.3$ & $22.03 \pm 0.86$ & $69.2 \pm 3.4$ \\
\hline \hline
\end{tabular}
\end{center}
\end{table}

We begin by identifying the algebraic source of information on $\alpha$ based
on known $B \to (\pi \pi,\rho \rho)$ rates and CP asymmetries.
The formalism for obtaining $\alpha$ from $B \to \pi \pi$ decays was proposed
in Ref.\ \cite{Gronau:1990ka} and is reviewed in Ref.\ \cite{Gronau:2007xg}.
One may define phases of amplitudes such that
\beq
A(B^0 \to \pi^+\pi^-) = |T|e^{i\gamma} + |P|e^{i\delta}~,
\eeq
where $|T|$ is the magnitude of a tree amplitude with weak phase $\gamma$,
while $|P|$ is the magnitude of a penguin amplitude with strong phase $\delta$.
The unitarity relation (\ref{eqn:ut}) has been used to express
$V^*_{tb}V_{td} = -V^*_{ub} V_{ud} - V^*_{cb} V_{cd}$, and
the resulting first term with a phase $\gamma$ incorporated into $T$.  An
initial $B^0$ or $\bar B^0$, defined by tagging the production vertex,
evolves as \cite{Gronau:1989ia,Gronau:2004ej}
\beq
\Gamma(B^0(t)/\bar B^0(t)) \sim e^{-\Gamma t}[1 \pm C_{+-} \cos \Delta m t
\mp S_{+-} \sin \Delta m t]
\eeq
with
\beq
C_{+-} \equiv \frac{1 - |\lambda_{\pi\pi}|^2}{1 + |\lambda_{\pi\pi}|^2}~,~~
S_{+-} \equiv \frac{2~{\rm Im}(\lambda_{\pi\pi})}{1 + |\lambda_{\pi\pi}|^2}~,~~
\lambda_{\pi\pi} \equiv e^{-2i\beta}\frac{A(\bar B^0 \to \pi^+\pi^-)}
{A(B^0 \to \pi^+\pi^-)}~.
\eeq
The tree transition $b \to u \bar u d$ carries isospin 1/2 and 3/2, while the
penguin transition $b \to d$ carries only isospin 1/2.  The spinless two-pion
state can only have isospin 0 and 2, so the $B \to \pi \pi$ amplitudes obey the
relation
\beq
A(B^0 \to \pi^+\pi^-)/\sqrt{2} + A(B^0 \to \pi^0 \pi^0) = A(B^+ \to \pi^+
\pi^0)~,
\eeq
with a corresponding relation for $\bar B$.  The amplitude $A(B^+ \to \pi^+
\pi^0)$ has no penguin contribution and thus has the weak phase $\gamma$, while
$A(B^- \to \pi^- \pi^0)$ has weak phase $-\gamma$.  Thus if we multiply all
$\bar B$ amplitudes by $e^{2i\gamma}$, defining them with a tilde, we can
express the triangle relations as
\beq
A_{+-}/\sqrt{2} + A_{00} = A_{+0}~,~~
\tilde A_{+-}/\sqrt{2} + \tilde A_{00} = \tilde A_{-0}~,
\eeq
where the triangles have the same base: $A_{+0} = \tilde A_{-0}$.  They would
be identical in the absence of the penguin amplitude, and recalling that
$\gamma + \beta = \pi - \alpha$, one would have $\sin(2\alpha) = S_{+-}/(1 -
C_{+-}^2)^{1/2}$.  The deviation from this value depends on the shapes of both
triangles, governed by the separate rates of $B$ and $\bar B$ decays.

The measurements used in our determination of $\alpha$ are summarized in
Table \ref{tab:measpp}.  They are taken from Ref.\ \cite{HFAG} except for
${\cal B}_{\rm av}(B^0 \to \pi^0 \pi^0)$, which is based on averaging a new
preliminary Belle measurement \cite{Vanhoefer:2015ksa} with an earlier BaBar
one (see Table \ref{tab:pzpz}), and $C_{00}$, which is taken from
Ref.\ \cite{PDG}.  The subscript ``av'' denotes the average for the process
and its CP conjugate.  We assume no CP violation in $B^+ \to \pi^+\pi^0$.  

\begin{table}
\caption{Inputs to the determination of $\alpha$ from an isospin analysis
of $B \to \pi \pi$ \cite{HFAG, PDG}.
\label{tab:measpp}}
\begin{center}
\begin{tabular}{c c c c} \hline \hline
Quantity & Value ($\times 10^{-6}$) & Quantity & Value \\ \hline
${\cal B}_{\rm av}(B^+ \to \pi^+ \pi^0)$ & $5.11 \pm 0.37^a$ 
 & $C_{+-}$ & $-0.31 \pm 0.05$ \\
${\cal B}_{\rm av}(B^0 \to \pi^+ \pi^-)$ & $5.12 \pm 0.19$
 & $C_{00}$ & $-0.43 \pm 0.24$ \\
${\cal B}_{\rm av}(B^0 \to \pi^0 \pi^0)$ & $1.17 \pm 0.13$
 & $S_{+-}$ & $-0.66 \pm 0.06$ \\ \hline \hline
\end{tabular}
\end{center}
\leftline{$^a$Branching ratio corrected by factor \cite{PDG} $\tau(B^0)/
\tau(B^+) = 0.929$.}
\end{table}

\begin{table}
\caption{Current (preliminary) status of $B_{\rm av}(B^0 \to \pi^0 \pi^0)$.
\label{tab:pzpz}}
\begin{center}
\begin{tabular}{c c} \hline \hline
Source & Value ($10^{-6}$)\\ \hline
Belle \cite{Vanhoefer:2015ksa} & $0.90 \pm 0.12 \pm 0.10$ \\
BaBar \cite{Lees:2012mma} & $1.83\pm0.21\pm0.13$ \\
Average & $1.165 \pm 0.132$ \\ \hline \hline
\end{tabular}
\end{center}
\end{table}

We obtain separate branching ratios for $B^0$ decays and their CP conjugates
using the relations
\beq
{\cal B}(B^0 \to f) = (1+C_f){\cal B}_{\rm av}(B^0 \to f)~,~~
{\cal B}(\bar B^0 \to f) = (1-C_f){\cal B}_{\rm av}(B^0 \to f)~.
\eeq
The sides of the triangles are then specified, and the angle $\theta_f$ between
the $B^0 (\bar B^0) \to f^0$ and $B^\pm \to f^\pm$ sides is calculated using
the law of cosines.  For $B\to\pi^+ \pi^-$ this yields $\theta_{+-} =
{\rm Arg}(A_{+-}/ A_{+0})$ for $B$ decays and $\tilde \theta_{+-} =
{\rm Arg}(\tilde A_{+-} /\tilde A_{-0})$ for $\bar B$ decays.  The
difference between these two angles, $\Delta \theta_{+-} = \tilde \theta_{+-}
- \theta_{+-}$, then may be used in the determination of $\alpha$ via the
relation 
\beq
\sin(2\alpha + \Delta\theta_{+-}) = \frac{S_{+-}}{\sqrt{1 - (C_{+-})^2}}~.
\eeq 
\begin{figure}
\begin{center}
\includegraphics[width=0.6\textwidth]{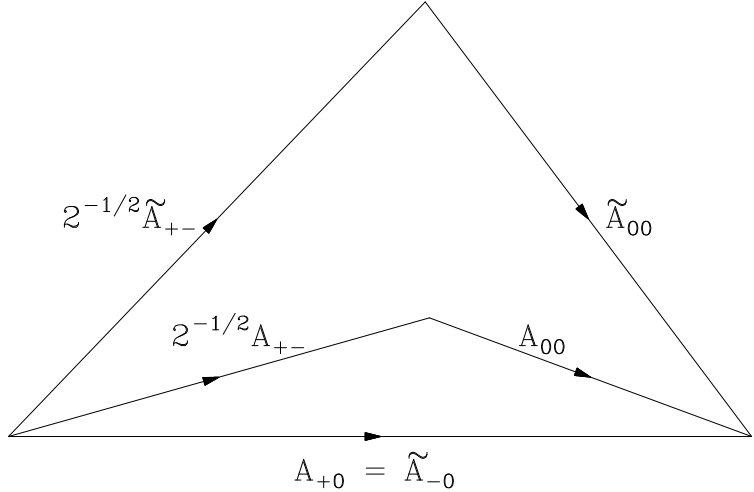}
\end{center}
\caption{Isospin triangles for the decays $B \to \pi \pi$.  Amplitudes for
$B$ decays are those without a tilde, while amplitudes with a tilde
correspond to those for $\bar B$ decays, multiplied by the phase
$e^{2 i \gamma}$ so that the bases of the two triangles coincide.
\label{fig:tri}}
\end{figure}
The triangles for a typical set of decays are shown in Fig.\ \ref{fig:tri}.
We shall also need the angles $\theta_{00}$ and $\tilde \theta_{00}$:
\beq
\theta_{00} \equiv {\rm Arg}(A_{00}/A_{+0})~,~~\tilde \theta_{00} \equiv
{\rm Arg}(\tilde A_{00}/\tilde A_{-0})~;~~\Delta \theta_{00} \equiv \tilde
\theta_{00} - \theta_{00}~,
\eeq 
determining the CP-violation parameter
\beq \label{eqn:s00def}
S_{00} = \sqrt{1 - (C_{00})^2}\sin(2\alpha + \Delta\theta_{00})~,
\eeq 
By definition, the $+-$ and $00$ angles have opposite signs.
Either triangle can be flipped about its base, giving a four-fold ambiguity
in $\Delta\theta_{+-}$ and hence $\alpha$. Furthermore, each value of $\sin
(2\alpha+\Delta \theta_{+-})$ corresponds to two values of $2\alpha + \Delta
\theta_{+-}$.  In practice \cite{CKMfitter,UTfit} all but one or two solutions
for $\alpha$ are incompatible with the unitarity relation (\ref{eqn:ut}).

We find solutions for $\alpha$ using a Monte Carlo program which generates the
six observables of Table \ref{tab:measpp} assuming they obey Gaussian
distributions.  One first generates the five observables $B_{+0}\equiv {\cal B}
(B^+ \to \pi^+ \pi^0)$, $B_{+-}^{\rm av} \equiv {\cal B}_{\rm av} (B^0 \to
\pi^+ \pi^-)$, $B^{\rm av}_{00} \equiv {\cal B}_{\rm av}(B^0 \to \pi^0 \pi^0)$,
$C_{+-}$, and $C_{00}$.  For the central values in Table \ref{tab:measpp}, the
$B$ triangle does not close, so the points of minimum $\chi^2 >0$ are those in
which it just barely closes, and hence lies flat with $\theta_{+-} = 0$.  The
contribution of the sixth observable $S_{+-}$ to $\chi^2$ depends on the
orientation of the isospin triangles through the quantity $\Delta \theta_{+-}$,
and the orientation giving the lowest $\chi^2$ is chosen.  (As $\theta_{+-}=0$
for the $B \to \pi \pi$ solutions with lowest $\chi^2$, only the sign of $\bar
\theta_{+-}$ matters.)  The predicted observables are updated each time a
Monte Carlo event gives a lower $\chi^2$ than found previously.  Typically one
obtains sufficient accuracy with 3 million generated events, though one must
smooth out fluctuations when isospin triangles are close to flat.
The values obtained are summarized in Table \ref{tab:fitpp}, with
individual $\chi^2$ contributions and their sum.

\begin{table}
\caption{Results of a fit to parameters determining $\alpha$ from an
isospin analysis of $B \to \pi \pi$.
\label{tab:fitpp}}
\begin{center}
\begin{tabular}{c c c c c c} \hline \hline
Quantity & Value ($\times 10^{-6}$) & $\chi^2$ & Quantity & Value & $\chi^2$ \\ \hline
${\cal B}_{\rm av}(B^+ \to \pi^+ \pi^0)$ & 5.019$^a$ & 0.061
 & $C_{+-}$ & $-0.303$ & 0.021 \\
${\cal B}_{\rm av}(B^0 \to \pi^+ \pi^-)$ & 5.134 & 0.006
 & $C_{00}$ & $-0.316$ & 0.227\\
${\cal B}_{\rm av}(B^0 \to \pi^0 \pi^0)$ & 1.190 & 0.023
 & $S_{+-}$ & $-0.66 \pm 0.06^b$ & \\
$\alpha$ (degrees) & $95.0,~141.1$ & \multicolumn{3}{c}{$\theta_{+-} =
 \theta_{00} = 0$}
 & $\chi^2_{\rm total} = 0.338$ \\
Other solutions & $128.9,~175.0$ & \multicolumn{3}{c}{$\tilde \theta_{+-} =
 33.9^\circ,~\tilde \theta_{00} = -54.6^\circ$} & \\ \hline \hline
\end{tabular}
\end{center}
\leftline{$^a$ Branching ratio corrected by factor \cite{PDG} $\tau(B^0)/
\tau(B^+) = 0.929$.}
\leftline{$^b$ Retained as input to determine $\alpha$.}
\end{table}

\begin{figure}
\begin{center}
\includegraphics[width=0.8\textwidth]{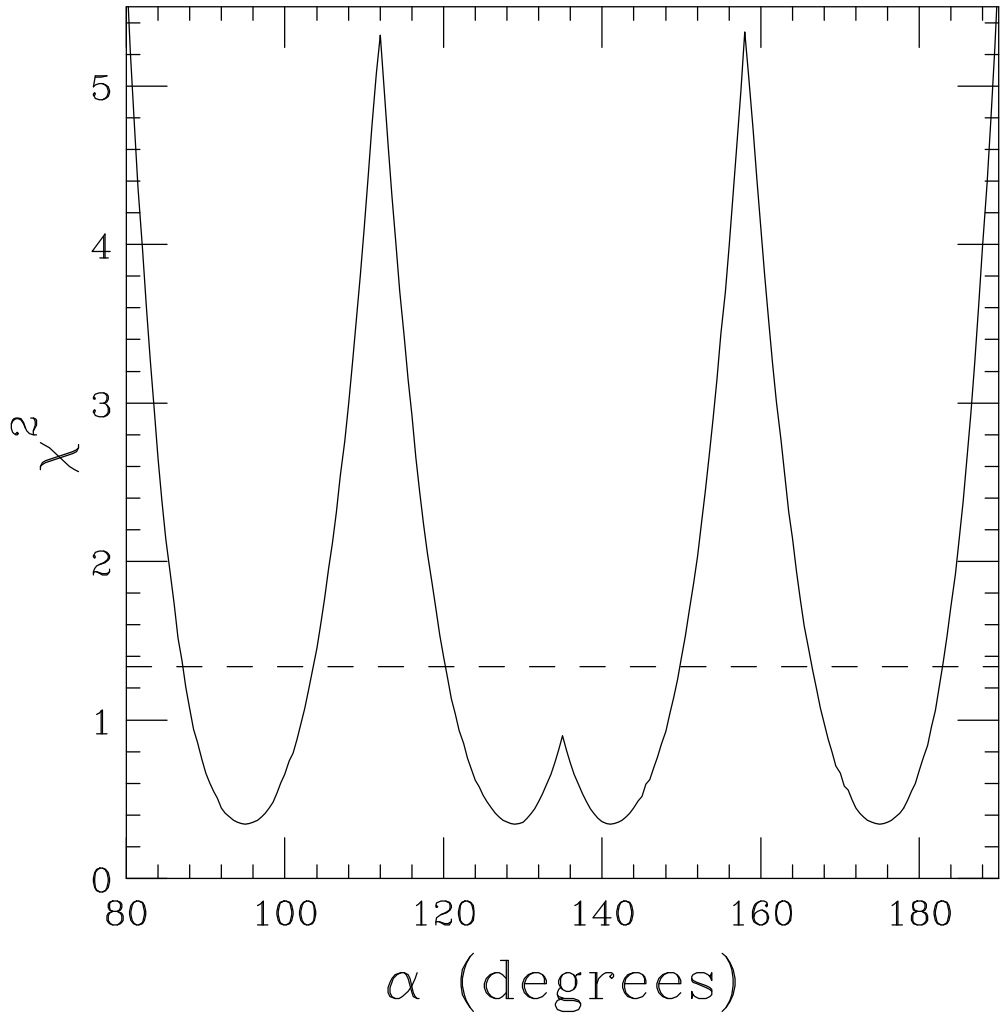}
\end{center}
\caption{Values of $\chi^2$ as a function of $\alpha = \phi_2$ as derived
from an isospin analysis of $B \to \pi \pi$.  The horizonal dashed line
denotes a value of $\chi^2$ one unit above the minimum.
\label{fig:fitpp}}
\end{figure}

The flatness of the $B$ isospin triangle in the favored fit means that the
eightfold ambiguity is reduced to a fourfold one, as only the $\bar B$ triangle
can be flipped.  A fit to the observables in Table \ref{tab:fitpp} results in
$\chi^2$ values shown in Fig.\ \ref{fig:fitpp}. [Fluctuations due to limited
Monte Carlo statistics have been smoothed out with piecewise parabolic fits
to regions near $\chi^2$ minima.]  Minimum values of $\chi^2 
= 0.338$ occur at $\alpha = (95,128.9,141.1,175)^\circ$.  $\Delta \chi^2 \le 1$
is satisfied for $\alpha$ in the range ([87,104],[120,150],[166,183])$^\circ$.
These results are in accord with those found by the CKMfitter Collaboration
\cite{CKMfitter}.  Note that for every solution $\alpha$, there is another
solution at $270^\circ - \alpha$, with both isospin triangles flipped so
that $\Delta \theta_{+-} \to -\Delta \theta_{+-}$.

\begin{figure}
\begin{center}
\includegraphics[width=0.92\textwidth]{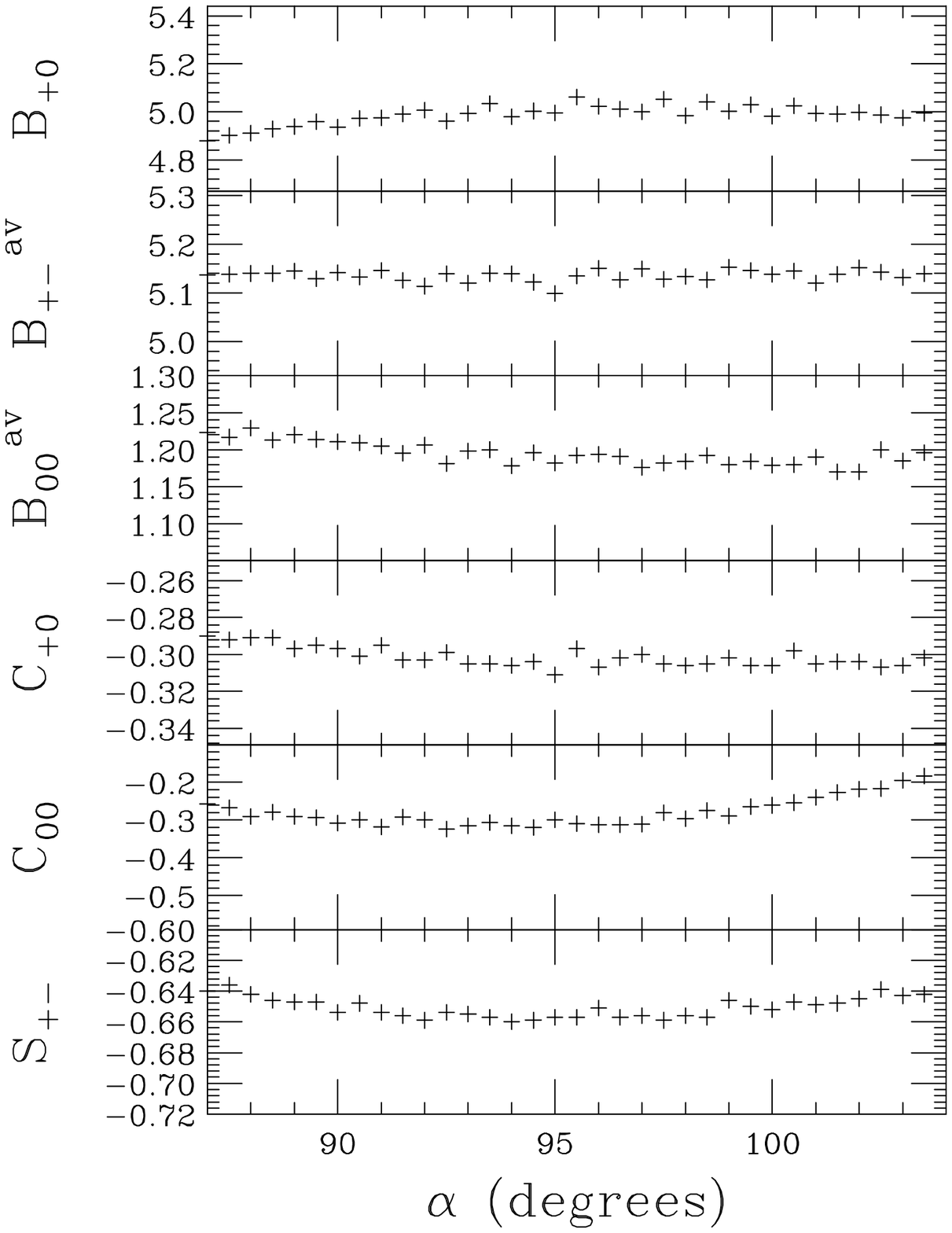}
\end{center}
\caption{Dependence of fitted input parameters describing $B \to \pi \pi$
decays on $\alpha$ in the range [87,104]$^\circ$.  Fluctuations are due to
limited Monte Carlo statistics.
\label{fig:vars}}
\end{figure}

In order to gauge the dependence of $\alpha$ on the input parameters, we
display their fitted values for the range $87^\circ \le \alpha \le 104^\circ$
in Fig.\ \ref{fig:vars}.

We note several features of the determination of $\alpha$ using only
$B \to \pi \pi$ decays.

\begin{itemize}

\item The greatest dependence of $\alpha$ on the parameters in Table
\ref{tab:measpp}, normalized by their experimental uncertainty, is on
$C_{00}$.  Indeed, the full $\pm 1 \sigma$ variation of $C_{00}$ is not
permitted.  If $C_{00}$ is too negative, the $B$ isospin triangle cannot close.
The requirement that the isospin triangles close was used in Ref.\
\cite{Gronau:2001ff} to place bounds on $B_{\rm av}(B^0 \to \pi^0 \pi^0)$
and on $\Delta\theta_{+-}$.

\item The uncertainty on ${\cal B}(B^+ \to \pi^+ \pi^0)$ has greater effect
on $\alpha$ than the experimental errors of either $B^0$ decay mode.

\item Reduction of $B_{\rm av}(B^0 \to \pi^0 \pi^0)$ reduces the allowable
parameter range for $C_{00}$, as it prevents the $B$ isospin triangle from
closing for a wider range of $C_{00}$.

\end{itemize}

The interplay of $C_{00}$ and $B_{\rm av}(B^0 \to \pi^0 \pi^0)$ is keenly
illustrated by the recent preliminary Belle value for the latter quantity
\cite{Vanhoefer:2015ksa}.  The significant reduction in $B_{\rm av}(B^0 \to
\pi^0 \pi^0)$ from the previous PDG average of $(1.91\pm0.22)\times 10^{-6}$
is what has prevented the $B$ isospin triangle from closing when all other
parameters are taken at their central values.  As stated in Ref.\
\cite{Vanhoefer:2015ksa}, any remeasurement of $B_{\rm av}(B^0 \to \pi^0\pi^0)$
must be regarded as preliminary until accompanied by a remeasurement of
$C_{00}$.

If a subsequent measurement finds $C_{00} = -0.316 \pm 0.12$, corresponding
to the fitted central value in Table \ref{tab:fitpp} with half the present
error while other inputs remain as in Table \ref{tab:measpp}, the
minimum $\chi^2$ is reduced to near zero, while the shape of the curve in
Fig.\ \ref{fig:fitpp} is essentially preserved.  Thus, the values of $\alpha$
at the minimum, and the range for which $\Delta \chi^2 < 1$, remain unchanged.

\begin{figure}
\includegraphics[width = 0.48\textwidth]{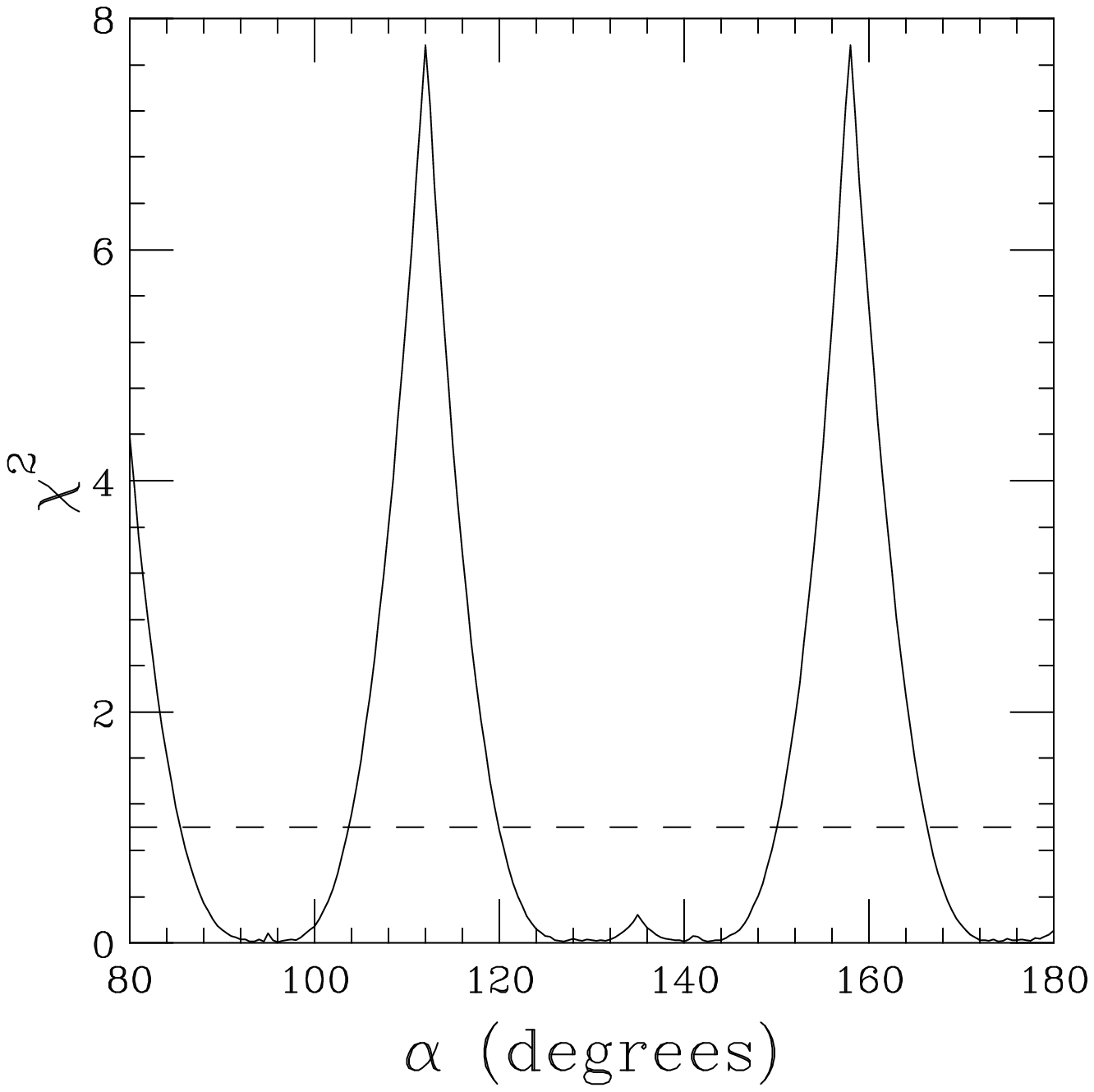}
\includegraphics[width = 0.48\textwidth]{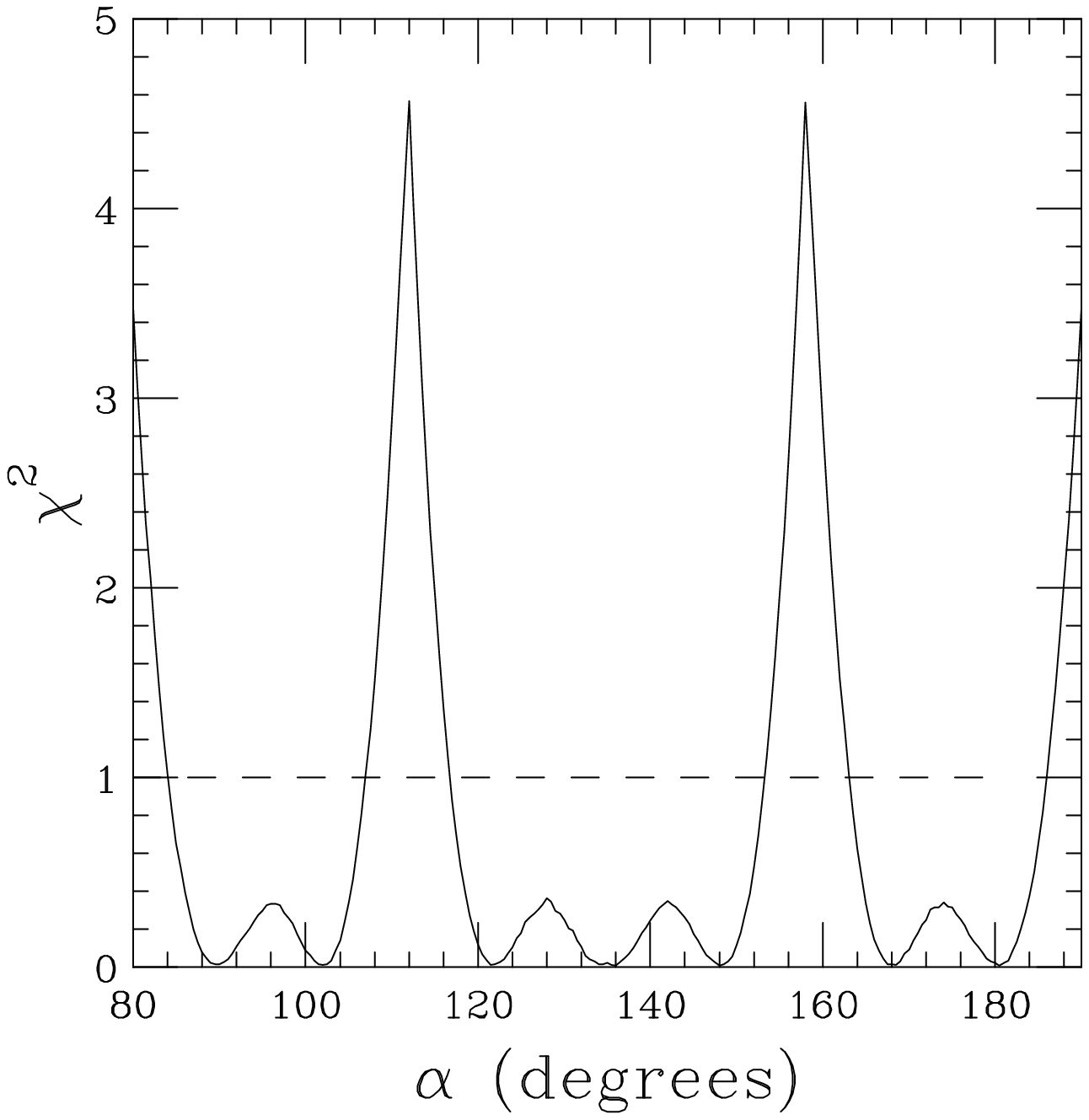}
\caption{Left: $\chi^2$ vs.\ $\alpha$ for central values of all parameters in
Table \ref{tab:fitpp}, with errors as in Table \ref{tab:measpp} except for
$\delta C_{00} = 0.12$.  Right:  Same except $C_{00} = -0.20 \pm 0.12$.
\label{fig:closed}}
\end{figure}

Now take central values of all parameters in Table \ref{tab:fitpp} with
errors as in Table \ref{tab:measpp} except for $\delta C_{00} = 0.12$.  The
resulting plot of $\chi^2$ vs.\ $\alpha$ is shown in the left-hand panel of
Fig.\ \ref{fig:closed}.  The $\chi^2$ curves are somewhat flattened at their
minima, but the values of $\alpha$ are not greatly affected.  If the central
value of $C_{00}$ is raised to $-0.2$, other parameters being kept fixed, the
resulting plot is shown in the right-hand panel of Fig.\ \ref{fig:closed}.
Here neither isospin triangle is flattened, so the full eight-fold degeneracy
of solutions occurs.  The $\chi^2$ minima are near 89.5, 102, 121.5, 134, 136,
148.5, 168, and 180.5 degrees (note the symmetry under $\alpha \leftrightarrow
270^\circ - \alpha$).  The ranges allowed for $\Delta \chi^2
\le 1$ are ([84,107],[117,153],[163,186])$^\circ$.  

Other parameters in Fig.\ \ref{fig:vars} which show some $\alpha$ dependence
are $B_{+0} \equiv {\cal B}(B^+ \to \pi^+ \pi^0)$ and $S_{+-}$.  We have
studied the effect of taking each parameter with half its present experimental
error.  The reduction of the error on $B_{+0}$ by a factor of two increases
the overall $\chi^2$ by less than 0.1.  Halving the $S_{+-}$ error reduces
the $\alpha$ range to ([88,103],[120,150],[167,182])$^\circ$.  Finally, the
effect of reducing all experimental errors in Table \ref{tab:measpp} by a
factor of two leads to an allowed $\alpha$ range of ([91,100],[124,146],
[170,179])$^\circ$.  Thus the error on $\alpha$ scales roughly as the error on
{\it all} six variables, while reducing the error on any individual variable
does not significantly affect the error on $\alpha$.

We next discuss the potential impact of a measurement of the time-dependent
CP-violation parameter $S_{00}$, given by Eq.\ (\ref{eqn:s00def}).
We may calculate $S_{00}$ for each orientation of the isospin triangles
and for each pair of $\alpha$ values resulting from the value of $\sin(2
\alpha + \Delta \theta_{00})$.  The results are shown in Table \ref{tab:alspp},
where the $B$ triangle has been taken to be flat.

\begin{table}[h]
\caption{Values of $\alpha$ consistent with the measurements in Table
\ref{tab:measpp}, and their corresponding values of $S_{00}$.  Angles are
given in degrees.  We are using $C_{00}$ from Table \ref{tab:fitpp}.
\label{tab:alspp}}
\begin{center}
\begin{tabular}{rrr} \hline \hline
                     & $\alpha$~~~~~ & $S_{00}$~~ \\ \hline
$\Delta \theta_{00} < 0$: &  $95.0^\circ$ & 0.67 \\
                   & or $141.1^\circ$ & --0.70 \\
$\Delta \theta_{00} > 0$: & $128.9^\circ$ & --0.70 \\
                   & or $175.0^\circ$ & 0.67 \\ \hline \hline
\end{tabular}
\end{center}
\end{table}

\begin{figure}[h]
\includegraphics[width=0.48\textwidth]{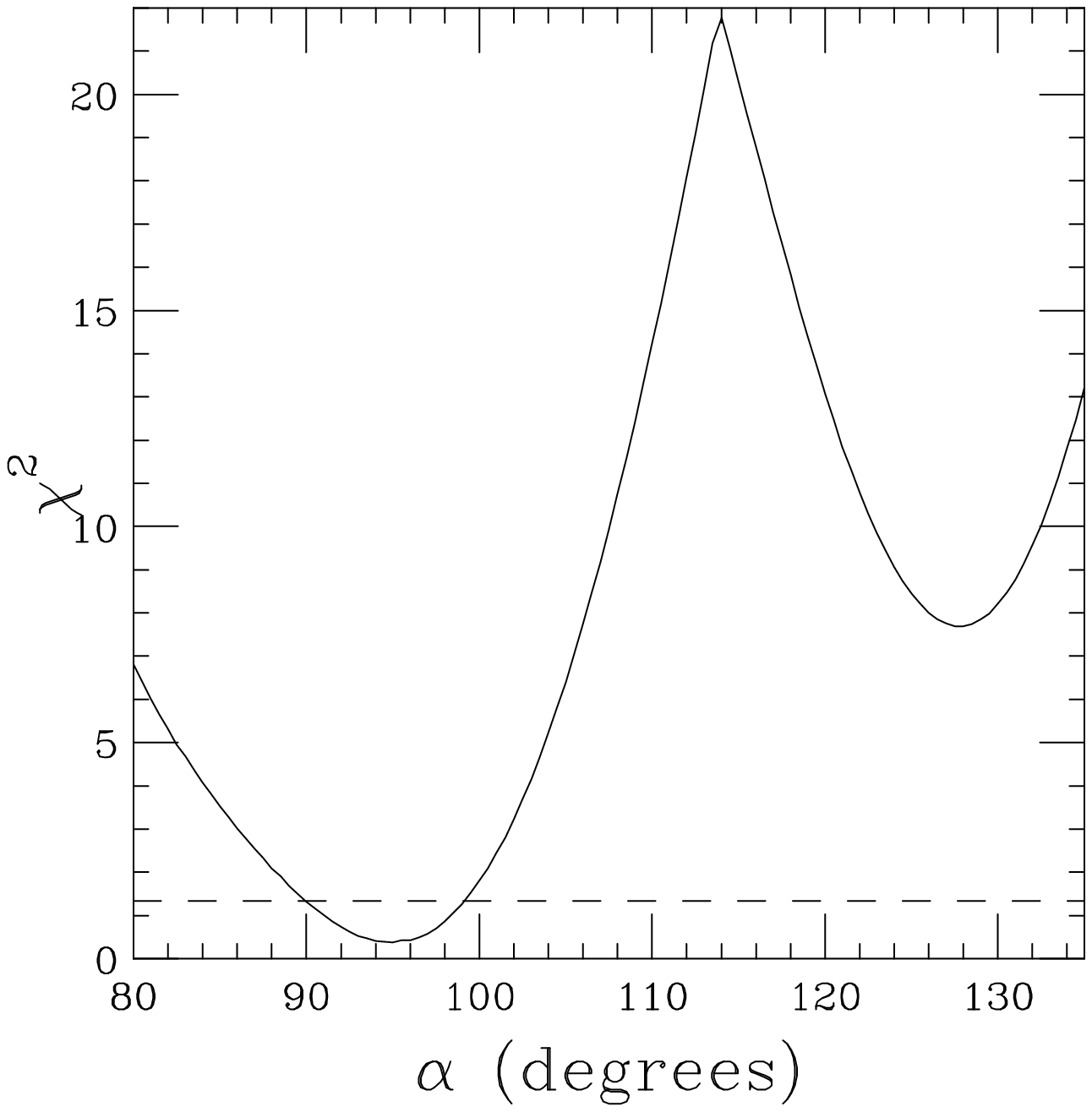}
\includegraphics[width=0.48\textwidth]{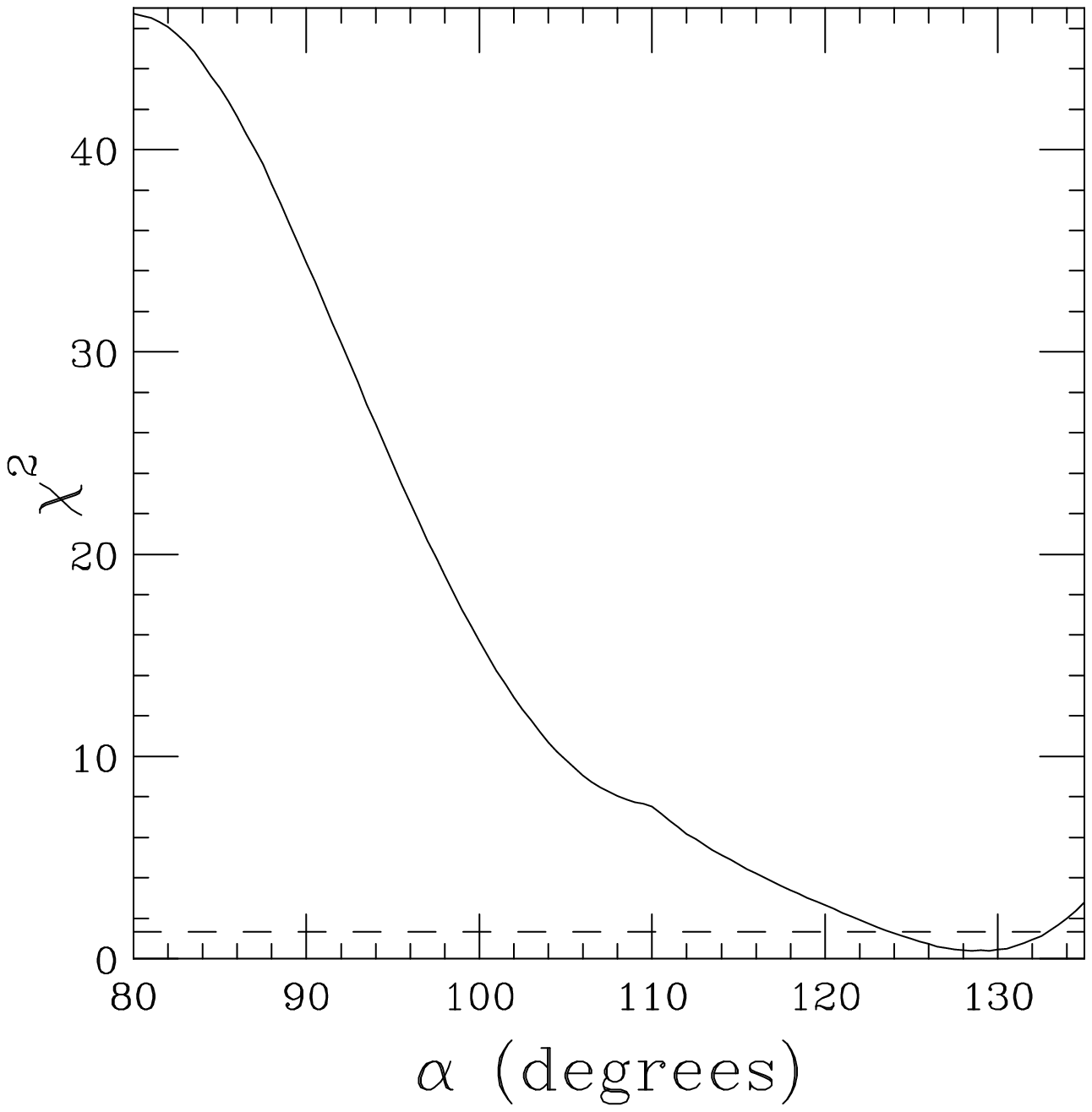}
\caption{Dependence of $\chi^2$ on $\alpha$ as extracted from an isospin
analysis of $B \to \pi \pi$ in the presence of a measurement of $S_{00}$.
Left: $S_{00} = 0.67 \pm 0.25$; right: $S_{00} = -0.70 \pm 0.25$.  We show only
the range $80^\circ \le\alpha \le 135^\circ$ because there exist solutions
with $\alpha \leftrightarrow 270^\circ - \alpha$ and the sign of $\Delta
\theta_{+-}$ changed.
\label{fig:s00}}
\end{figure}

Future measurements of $S_{00}$ at the Belle II $B$ factory using external
photon conversion on a data sample of $50\times 10^9$ $B\bar B$ pairs
\cite{Ishino:2007pt} may be able to favor one of the two predicted values of
$S_{00}$ over the other.  As an example, we compare in Fig.\ \ref{fig:s00}
the $\chi^2$ dependence on $\alpha$ when $S_{00} = 0.67 \pm 0.25$ (left)
or $-0.70 \pm 0.25$ (right).

A distinction between solutions with $\alpha = (95,175)^\circ$ and $(129,141)
^\circ$ is possible.  The allowed ranges of $\alpha$ within these solutions
are reduced slightly (e.g., to the interval [90,99] degrees). There still
remains a two-fold ambiguity in $\alpha$.  Anticipating a value of $\alpha$
near $90^\circ$ consistent with other CKM constraints, the second solution near 
$180^\circ$ with the same value of $S_{00}$ can be then easily excluded.

We now perform similar analyses for $B \to \rho \rho$ decays.  We use
branching fractions multiplied by the fraction $f_L$ of decays leading to
longitudinal $\rho$ polarization.  We first examine inputs analogous to the
six $B \to \pi \pi$ observables: three $B$'s, two $C$'s, and $S_{+-}$.
They are listed in Table \ref{tab:measrr}.  The inputs leading to the first
three entries are summarized in Table \ref{tab:flb}.

\begin{figure}
\includegraphics[width=0.48\textwidth]{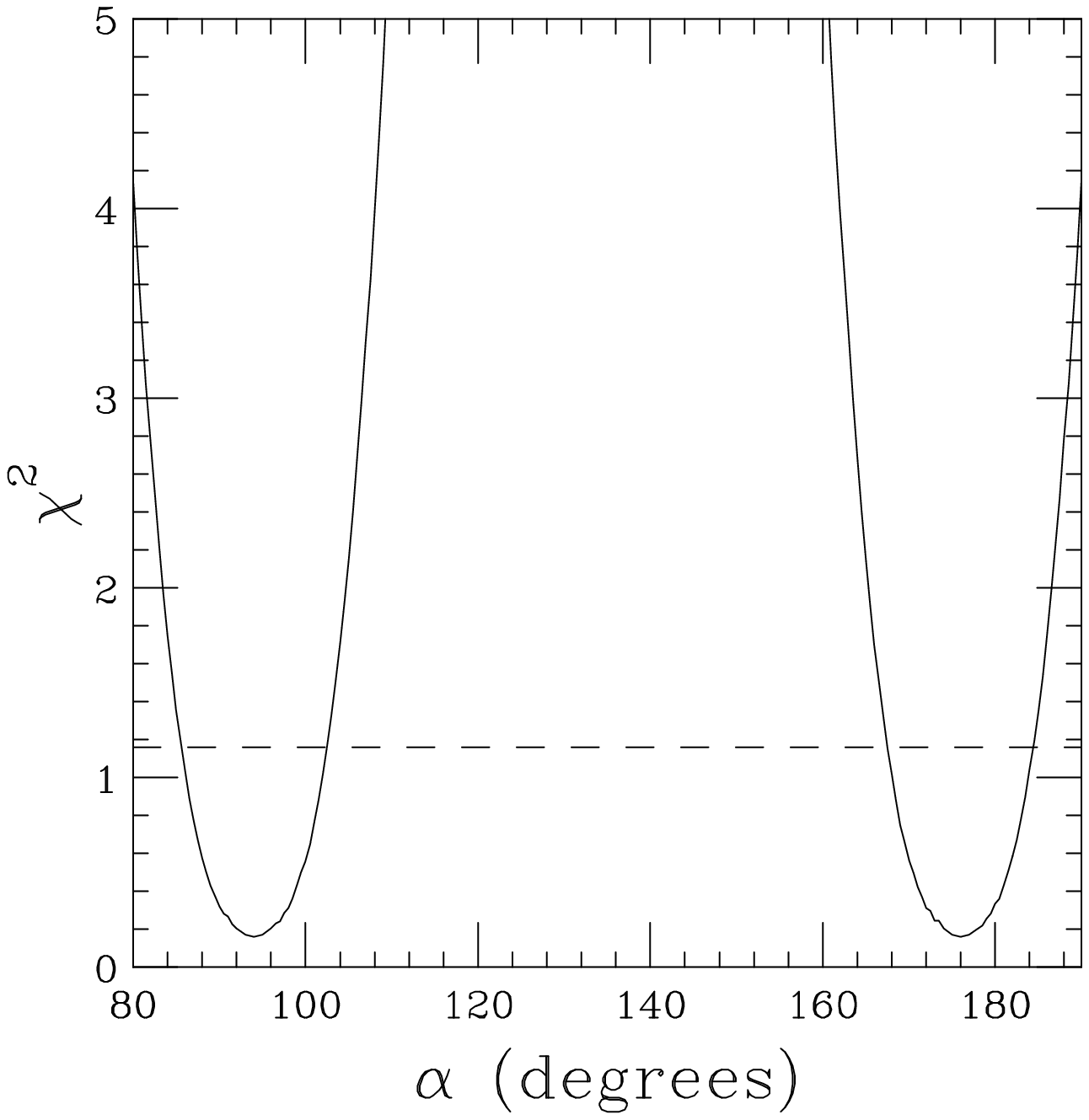}
\includegraphics[width=0.48\textwidth]{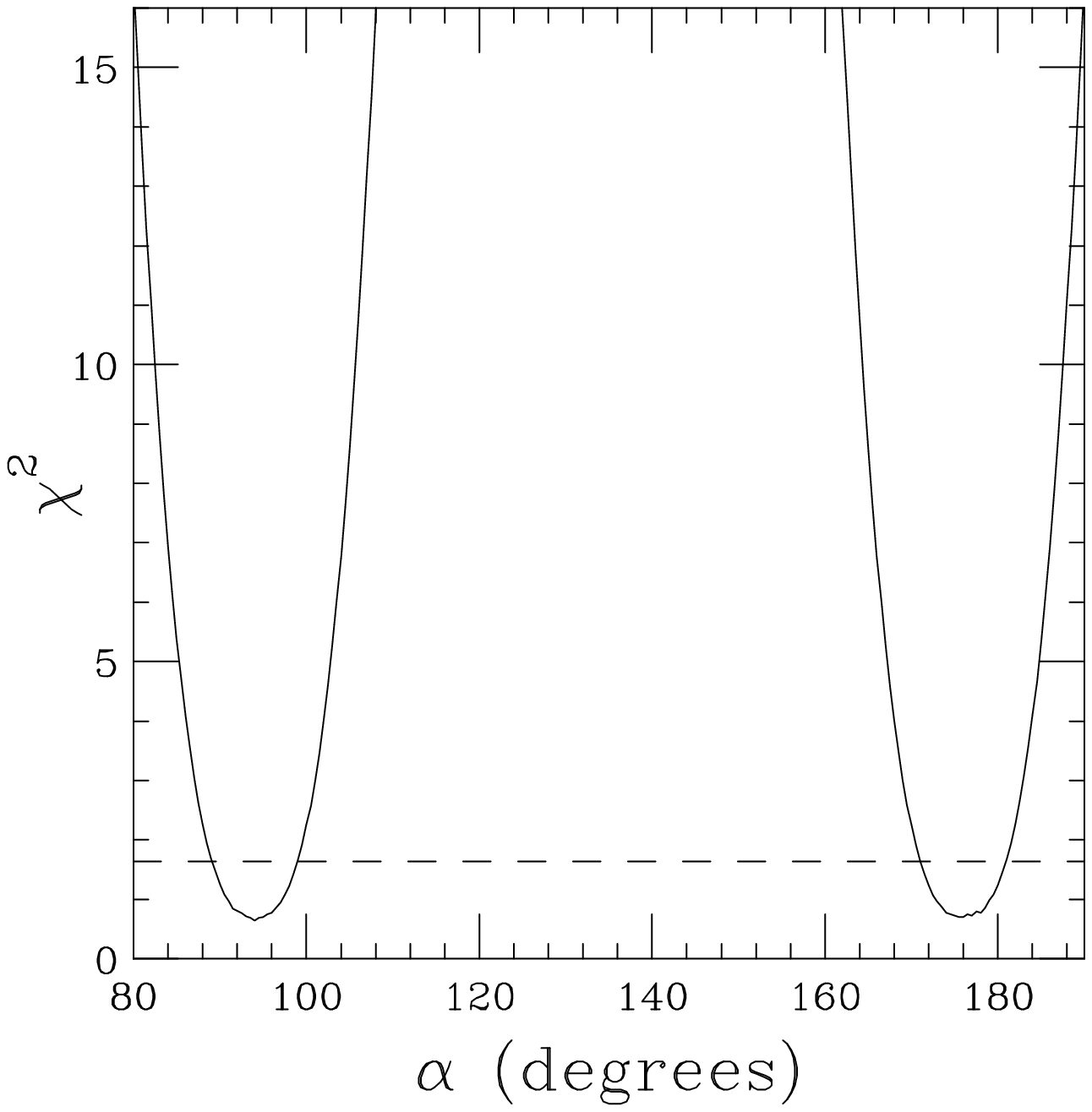}
\caption{Values of $\chi^2$ as a function of $\alpha = \phi_2$ from an isospin
analysis of $B \to \rho \rho$ based on the six parameters of Table
\ref{tab:measrr}.  Left:  present experimental errors, with $\Delta \chi^2 \le
1$ corresponding to $\alpha = (94 \pm 8)^\circ$ or $(176 \pm 8)^\circ$.
 Right: present errors divided by two, leading to $\alpha = (94 \pm 5)^\circ$
or $(176 \pm 5)^\circ$.
\label{fig:rr}}
\end{figure}

Here, both the $B$ and $\bar B$
triangles fail to close for the listed central values.  A $\chi^2$ fit to the
first five parameters yields the values in Table \ref{tab:fitrr}. As in the
case of $B \to \pi \pi$, these parameters are those which make the $B$ triangle
exactly flat.  In this case the $\bar B$ triangle is also flat, leading to a
degeneracy of solutions.  The $\chi^2$ distributions for nominal variables and
for the same central values with errors divided by two are shown in Fig.\
\ref{fig:rr}.

\begin{table}
\caption{Inputs to the determination of $\alpha$ from an isospin analysis
of $B \to \rho \rho$.  Observed branching fractions are multiplied by
observed longitudinal $\rho$ polarization fractions \cite{HFAG, PDG}.
\label{tab:measrr}}
\begin{center}
\begin{tabular}{c c c c} \hline \hline
Quantity & Value ($\times 10^{-6}$) & Quantity & Value \\ \hline
$f_L{\cal B}_{\rm av}(B^+ \to \rho^+ \rho^0)$ & $21.18 \pm 1.71^a$
 & $C_{+-}$ & $0.00 \pm 0.09$ \\
$f_L{\cal B}_{\rm av}(B^0 \to \rho^+ \rho^-)$ & $27.42 \pm 1.95$
 & $C_{00}$ & $0.20 \pm 0.85$ \\
$f_L{\cal B}_{\rm av}(B^0 \to \rho^0 \rho^0)$ & $0.67 \pm 0.12^b$
 & $S_{+-}$ & $-0.14 \pm 0.13$ \\ \hline \hline
\end{tabular}
\end{center}
\leftline{$^a$Branching ratio corrected by factor \cite{PDG} $\tau(B^0)/
\tau(B^+) = 0.929$.}
\leftline{$^b$Averaged values of branching ratio and longitudinal fraction 
using also Ref.\ \cite{Aaij:2015ria}.}
\end{table}

\begin{table}
\caption{Individual measurements used to calculate longitudinal branching
fractions (first three entries of Table \ref{tab:measrr}).  We denote
${\cal B}^{ij} \equiv {\cal B}(B \to \rho^i \rho^j)$ given in units of
$10^{-6}$, $f_L^{ij} \equiv f_L(B \to \rho^i \rho^j)$.
\label{tab:flb}}
\begin{center}
\begin{tabular}{c c c c c} \hline \hline
Quantity & Belle\cite{Zhang:2003up,Vanhoefer:2015ijw,Adachi:2012cz}
 & Babar \cite{Aubert:2009it,Aubert:2007nua,Aubert:2008au}
 & LHCb \cite{Aaij:2015ria} & Average   \\ 
\hline
${\cal B}^{+0}$ & 31.7$\pm$7.1$^{+3.8}_{-6.7}$ 
& 23.7$\pm$1.4$\pm$1.4 & -- & 24.0$\pm$1.9 \\
$f_L^{+0}$ & 0.95$\pm$0.11$\pm$0.02 & 
0.950$\pm$0.015$\pm$0.006 & -- & 0.950$\pm$0.016\\
${\cal B}^{+-}$ & 28.3$\pm$1.5$\pm$1.5 & 
25.5$\pm$2.1$^{+3.6}_{-3.9}$ & -- & 27.7$\pm$1.9 \\
 $f_L^{+-}$ & 0.988$\pm$0.012$\pm$0.023 & 
 0.992$\pm$0.024$^{+0.026}_{-0.013}$ & -- & 0.990$\pm$0.019 \\
 ${\cal B}^{00}$ & 1.02$\pm$0.30$\pm$0.15 &
 0.92$\pm$0.32$\pm$0.14 & 0.94$\pm$0.17$\pm$0.09$\pm$0.06 & 0.95$\pm$0.15 \\
 $f_L^{00}$ & 0.21$^{+0.18}_{-0.22}\pm$0.15 & 0.75$^{+0.11}_{-0.14}\pm$0.05
 & 0.745$^{+0.048}_{-0.058}\pm$0.034 & 0.71$\pm$0.06 \\
\hline \hline
\end{tabular}
\end{center}
\end{table}

\begin{table}
\caption{Results of a fit to the six parameters in Table \ref{tab:measrr}.
\label{tab:fitrr}}
\begin{center}
\begin{tabular}{c c c c c c} \hline \hline
Quantity & Value ($\times 10^{-6}$) & $\chi^2$ & Quantity & Value & $\chi^2$ 
 \\ \hline
$f_L{\cal B}_{\rm av}(B^+ \to \rho^+ \rho^0)$ & 20.73$^a$ & 0.070
 & $C_{+-}$ & $-0.008$ & 0.008 \\
$f_L{\cal B}_{\rm av}(B^0 \to \rho^+ \rho^-)$ & 27.78 & 0.034 &
 $C_{00}$ & 0.036 & 0.037 \\
${f_L\cal B}_{\rm av}(B^0 \to \rho^0 \rho^0)$ & 0.68 & 0.011
 & $S_{+-}$ & $-0.14\pm0.13^b$ & \\
$\alpha$ (degrees) & $94,~176$ & & & $\chi^2_{\rm total}$ & 0.160 \\
\hline \hline
\end{tabular}
\end{center}
\leftline{$^a$ Branching ratio corrected by factor \cite{PDG} $\tau(B^0)/
\tau(B^+) = 0.929$.}
\leftline{$^b$ Retained as input to determine $\alpha$.}
\end{table}

The greatest sensitivity of $\alpha$ to the measurements in Table
\ref{tab:measrr}, normalized by their experimental uncertainty, originates in
$f_L{\cal B}_{\rm av}(B^+ \to \rho^+ \rho^0)$ and $f_L{\cal B}_{\rm av}(B^0
\to \rho^+ \rho^-)$.  More precise information on branching fractions
would be helpful.
Significant improvement is expected in thirteen-year-old Belle results for 
$B^+ \to \rho^+\rho^0$~\cite{Zhang:2003up}, based on only about ten 
percent of the final Belle $\Upsilon(4S)$ sample.

An additional piece of experimental information is available in the case of
$B \to \rho \rho$.  The BaBar Collaboration \cite{Aubert:2008au}
has measured
\beq \label{eqn:s00}
S_{00} = 0.3 \pm 0.7 \pm 0.2 = 0.3 \pm 0.73~.
\eeq
Despite its large uncertainty, this measurement has a significant
effect on $\alpha$.  There are now two quantities, $S_{+-}$ and $S_{00}$,
which depend on $\alpha$.  With $S_{+-}$ alone, a $\chi^2$ fit is governed
solely by the geometry of the isospin triangles.  When both $S_{+-}$ and
$S_{00}$ are specified, some tension can arise between their favored values of
$\alpha$, and the geometry of the isospin triangles can be adjusted to minimize
this tension.

\begin{figure}
\includegraphics[width=0.32\textwidth]{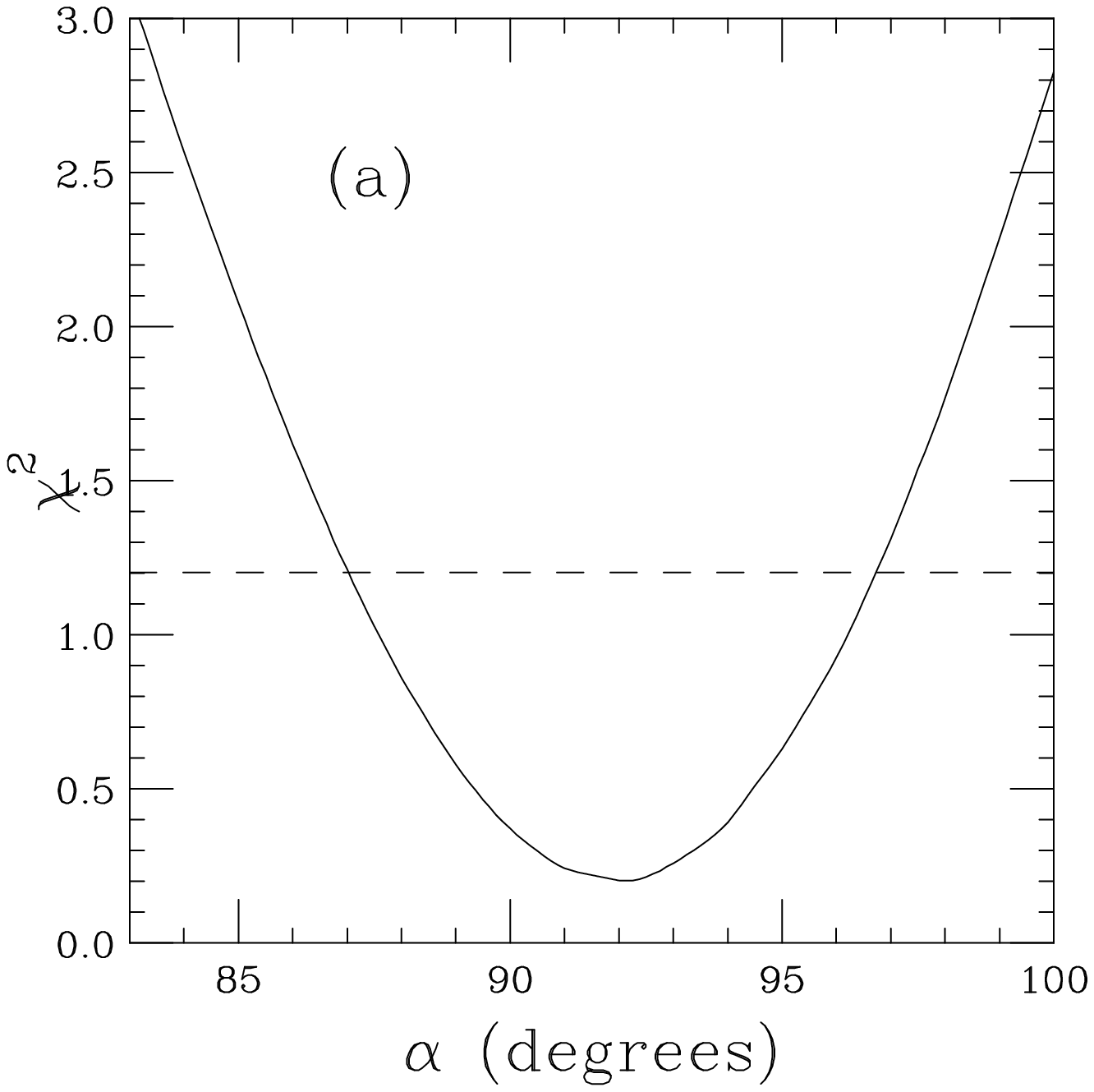}
\includegraphics[width=0.32\textwidth]{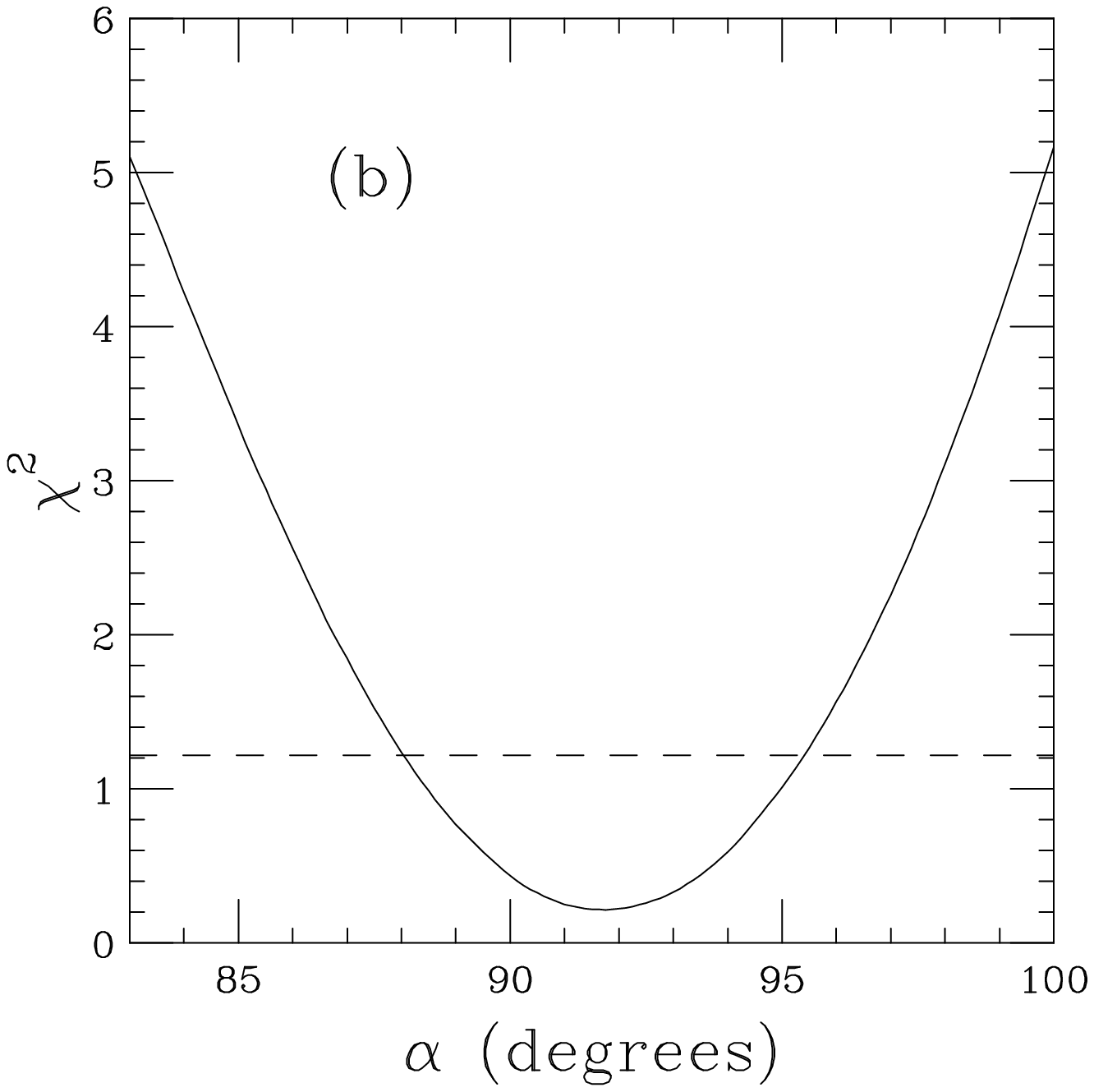}
\includegraphics[width=0.32\textwidth]{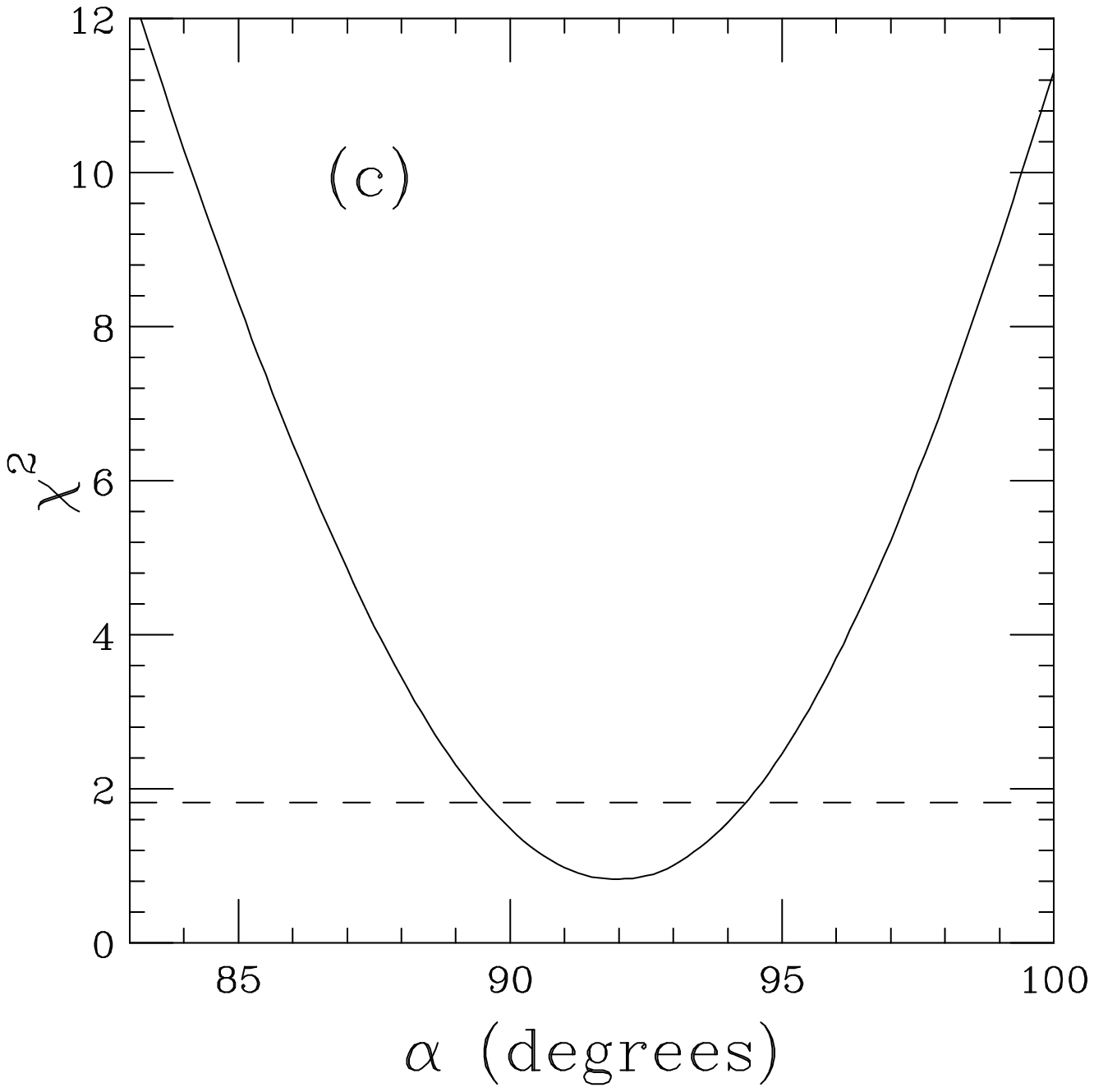}
\caption{Isospin triangle fits to $B \to \rho \rho$ observables in Table
\ref{tab:measrr} when the measurement (\ref{eqn:s00}) is included.  (a)
nominal experimental errors.  (b) Same as (a) but with present error on
$S_{00}$ divided by two.  (c) Same as (a) but with {\it all} experimental
errors divided by two.
\label{fig:s00rr}}
\end{figure}

We show in Fig.\ \ref{fig:s00rr}(a) the effect of adding the observable
(\ref{eqn:s00}), related to $\alpha$ through Eq.\ (\ref{eqn:s00def}), to those
in Table \ref{tab:measrr}. (We show only the solution consistent with other
observables.)  The value of $\alpha$ corresponding to $\Delta \chi^2 \le 1$ is
now $(92.0^{+4.7}_{-5.0})^\circ$.  In Fig.\ \ref{fig:s00rr}(b)  we show the
$\chi^2$ distribution when the error on $S_{00}$ is divided by two, leading to
$\alpha = (91.7^{+3.8}_{-3.7})^\circ$.  
We also checked that a substantial reduction of the error on
$f_LB_{\rm av}(B^0 \to \rho^0 \rho^0)$, potentially achievable at the LHCb 
upgrade, would have an insignificant effect on improving the precision in $\alpha$.
Finally, in Fig.\ \ref{fig:s00rr}(c) we show the $\chi^2$ distribution when {\it all} 
errors are divided by two, in which case one finds $\Delta \chi^2 \le 1$ for 
$\alpha = (92.0 \pm 2.5)^\circ$.

We have discussed ways to narrow the uncertainty in the CKM phase $\alpha =
\phi_2$ as derived from isospin analyses of $B \to \pi \pi$ and $B \to \rho
\rho$.  No single variable in $B \to \pi \pi$ dominates the present error of
$9^\circ$ in $\alpha$.  Reduction of that error by a factor of two {\it is}
achieved if the errors in all six inputs of Table \ref{tab:measpp} are cut in
half.  The time-dependent CP violation parameter $S_{00}$ will help to
distinguish solutions near $\alpha = 129^\circ$ and $141^\circ$, yielding
$S_{00} \simeq -0.70$, from those near $95^\circ$ and $175^\circ$, yielding
$S_{00} \simeq 0.67$.

For the $B \to \rho \rho$ analysis, improving measurements of longitudinal
branching fractions of $B^+\to \rho^+\rho^0$ and $B^0\to \rho^+\rho^-$ would
reduce the $5^\circ$ current error in $\alpha$ as determined in these
processes.  The measurement of $S_{00}$ in $B^0 \to \rho^0 \rho^0$ with an
error reduced by a factor of two (or more) also would have a significant effect
on the accuracy of determining $\alpha$.  However, reduction by a factor of two
of {\it all} experimental errors (including that of $S_{00}$)
would reduce the error on $\alpha$ to $2.5^\circ$, a point at which one should
begin to take into account the $\rho$ width and isospin-breaking corrections.

\bigskip

We thank Shlomo Dado, Mark Oreglia, and Yoram Rozen for helpful discussions.
J.L.R. is grateful to the Technion for its generous hospitality during the
inception of this work, which was supported in part by the United States
Department of Energy through Grant No.\ DE-FG02-13ER41598.  He also thanks the
Mainz Institute for Theoretical Physics (MITP), the Universit\`a di Napoli
Federico II, and INFN for its hospitality and its partial support during a
portion of this work.

\end{document}